\documentclass[
  reprint,
  amsmath,
  amssymb,
  aps,
  prl,
  longbibliography,
  superscriptaddress,
]{revtex4-2}

\usepackage{graphicx}
\usepackage[colorlinks=true,linkcolor=blue,citecolor=blue,urlcolor=blue]{hyperref}

\graphicspath{{figures/}}

\begin{document}
\title{Size-Dependent Growth Rates Amplify Infinitesimal Asymmetry in Nanocrystals}
\author{Sam Oaks-Leaf}
\affiliation{Department of Chemistry, University of California, Berkeley, Berkeley, CA, 94720 USA}
\affiliation{Materials Sciences Division, Lawrence Berkeley National Laboratory, Berkeley, CA, USA}
\author{David T. Limmer}
\email{dlimmer@berkeley.edu}
\affiliation{Department of Chemistry, University of California, Berkeley, Berkeley, CA, 94720 USA }
\affiliation{Chemical Sciences  Divisions, Lawrence Berkeley National Laboratory, Berkeley, CA, 94720 USA}
\affiliation{Materials Sciences Division, Lawrence Berkeley National Laboratory, Berkeley, CA, USA}
\affiliation{Kavli Energy NanoSciences Institute, Berkeley, CA, 94720 USA}
\date{\today}

\date{\today}

\begin{abstract}
The kinetic Wulff construction predicts symmetric nonequilibrium shapes when crystallographically equivalent facets share a fixed growth rate. However, nanocrystals grow through finite facets whose nucleation barriers and ligand coverages depend on facet size. Here we develop a size-dependent kinetic construction and show that as a consequence infinitesimal seed asymmetries can be amplified into strongly asymmetric nanocrystal morphologies even when all symmetry-related facets obey the same microscopic growth law. Spatially heterogeneous nucleation rates and boundary limited growth generate facet growth velocities that depend on size, and deterministic shape evolution
translates these local rates into global shape symmetry breaking. We illustrate the mechanism of persistent anisotropic growth on square and triangular lattices in two dimensions and on FCC seeds in three dimensions, where cuboctahedra can evolve toward rods or tetrahedra  depending on which facet-area perturbations are amplified.
\end{abstract}

\maketitle

The ability to grow nanocrystals with prescribed shapes is central to using colloidal materials as optical, catalytic, mechanical, and self-assembling building blocks. Small changes in shape can change the exposed facet area, surface reactivity, plasmonic response, and packing behavior of nanoparticles \cite{yang2019anisotropic,cozzoli2006synthesis,scher2003shape,henzie2012self}. As a result, shape-controlled synthesis has developed a rich empirical toolbox based on seeds, ligands, solution conditions, and reaction
time~\cite{grzelczak2008shape,murphy2011gold,xia_shape_2015,yang_surface_2020}. A predictive theory of nanocrystal shape must explain not only which crystallographic facets
are stabilized \cite{marks2016nanoparticle}, but also why initially similar seeds can diverge toward rods, plates, tetrahedra, or other lower-symmetry products. Such a theory that translates microscopic kinetics into predictions of asymmetric growth is currently lacking. Building off of recent work on simple models \cite{oaks_leaf_mechanism_2025}, we introduce a theory demonstrating that shape symmetry breaking originates generically from the size-dependence of facet growth velocities.

The usual theoretical starting point for nanocrystal growth is the Wulff construction and its kinetic counterpart \cite{boukouvala2021approaches}. In equilibrium, the Wulff construction prescribes facet distances set by surface free energies resulting in compact structures with low total surface area and close-packed crystalline facets.
Under irreversible growth, the kinetic Wulff construction replaces surface energy with a crystallographic growth velocity. Under that construction, fast-growing facets disappear, whereas slow-growing facets bound the observed
morphology~\cite{marks_nanoparticle_2016,xia_kinetic_2025,balankura_predicting_2016,fichthorn_multi_2016}.
These models are useful because they compress complicated microscopic chemistry into a small number of parameters. However, both the equilibrium and kinetic constructions have a built-in symmetry constraint. All facets related by crystal symmetry are assigned the same properties, and consequently the construction preserves that symmetry upon growth. Cubes, octahedra, and related polyhedra can be selected by changing surface free energy or velocity ratios, but equivalent facets do not spontaneously break symmetry.

Experiments and simulations show that this symmetry-preserving picture is incomplete. Seed-mediated syntheses routinely access strongly anisotropic or lower-symmetry morphologies, including metal nanorods and tetrahedral
nanocrystals grown from higher-symmetry precursors~\cite{jana2001wet,dumestre2002shape,sun2021understanding,nguyen_colloidal_2022}. Recent theory has clarified two important pieces of this problem. Bassani and Engel used rejection-free kinetic Monte Carlo simulations to show how terrace-ledge-kink kinetics and growth-island geometry can kinetically trap FCC nanocrystals in a broad family of symmetry-preserving shapes~\cite{bassani2025kinetically}. In a complementary direction, we have shown that surfactant coverage fluctuations can be amplified by a positive feedback between adsorbate density and facet growth, providing a route from symmetric initial conditions to asymmetric products~\cite{oaks_leaf_mechanism_2025}.

We propose that finitely sized facets are the key ingredient to predict shape symmetry breaking in nanocrystal growth. Layer completion on a nanocrystal facet is not the same process as growth on a macroscopic surface \cite{levi1997theory,misbah2010crystal}.
The boundary of a finite facet limits the growth of a  nucleus originating in the interior of the facet, and adsorbates or surfactants can make nucleation rates different at edges, corners, and facet interiors \cite{lagrow2019situ,lee2026nanocrystal,li2021corner}. The result is a growth velocity that generically depends on facet size \cite{widmer2016ligand} not only on crystallographic orientation. With size dependent growth velocities, equivalent facets with infinitesimally different sizes can grow at different instantaneous rates even though they obey the same microscopic law. In this sense, size-dependent growth generalizes the kinetic Wulff construction from a static assignment of velocities in each crystallographic direction into a set of coupled dynamical equations where the current shape controls the nanocrystal's own subsequent growth. The dynamical feedback that emerges can sustain persistent asymmetric growth from infinitesimal initial perturbations.  

Here we develop this idea in three geometries of increasing complexity. First, on the square lattice, a length-dependent velocity can amplify an area-preserving perturbation into rod-like growth. Second, on the triangular lattice, the same kind of velocity law produces qualitatively different outcomes because advancing a close-packed facet shortens that facet. Surface-area-limited growth converts small perturbations into triangles, rhombi, or rod-like structures. Finally, for FCC nanocrystals bounded by $\{100\}$ and $\{111\}$ facets, we find that cuboctahedra seeds can grow into rods or symmetry lowering tetrahedra from infinitesimal facet area perturbations. Across these examples, we show that strong shape asymmetry does not require explicitly anisotropic chemistry among symmetry-equivalent facets. It can emerge from the coupling between finite-size facet growth rates and the evolving geometry of the crystal itself.

\begin{figure*}[ht]
    \centering
    \includegraphics[width=16cm]{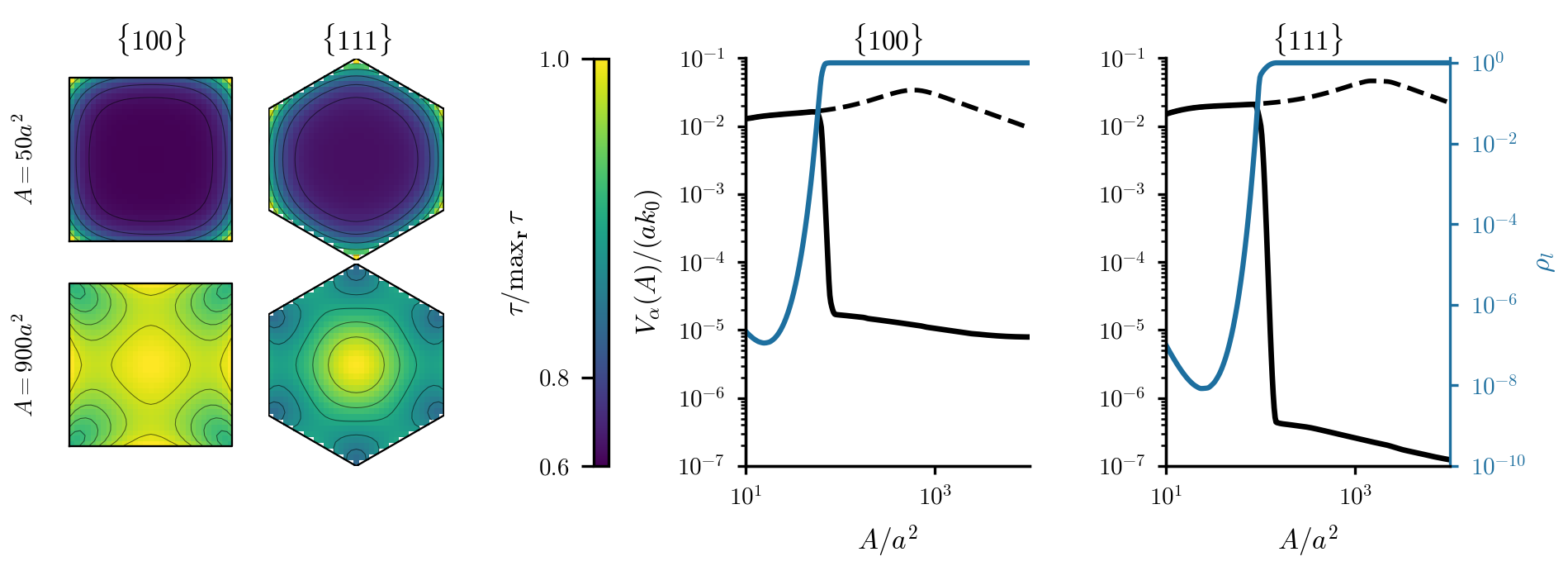}
    \caption{Microscopic origin of facet-area-dependent growth velocities.
    Left, normalized layer-completion times,
    $\tau/\mathrm{max}_{\mathbf{r}}\tau$, for finite square $\{100\}$ and
    hexagonal $\{111\}$ facets with $A=50a^2$ and $900a^2$ using corner, edge,
    and interior nucleation rates obtained from the ligand-renormalized
    microscopic attachment/removal rates. Right, dimensionless velocities
    $V_\alpha(A)/(ak_0)$ for the same facet families. Solid black curves
    include ligand blocking, dashed black curves show the corresponding
    ligand-free velocities, and blue curves show the mean ligand density
    $\rho_l$. In all panels $a=1$, $k_0=1$, $\beta\epsilon_n=5$, and
    $\beta\epsilon_l=4$. For $\{100\}$ facets,
    $z_{100}=2$, $\beta\Delta\mu_n=7$, and
    $\beta\Delta\mu_l=1$. For $\{111\}$ facets, $z_{111}=3$,
    $\beta\Delta\mu_n=12$, and
    $\beta\Delta\mu_l=0.744$.}
    \label{fig:size-dependent-growth-laws}
\end{figure*}

The mechanism proposed here for spontaneous shape symmetry breaking requires a facet velocity that changes with the
instantaneous facet size. A  facet advances when a new layer
nucleates and then spreads laterally. On an infinite surface this process can be summarized by a single nucleation rate and step speed, but on a nanocrystal  the layer must finish within a bounded polygon. The boundary truncates the set of possible nucleation events that can cover a point and makes nucleation sites at corners, edges, and interiors of the facet distinct. Neverthesless, we can describe the time to grow a new layer with the classic probabilistic model due to Avrami, Kolmogorov, Johnson, and Mehl~\cite{avrami1940kinetics,kolmogorov1937statistical,johnson1939mehl} and later modified to account for finite domains by Cahn~\cite{cahn_time_1995}. 

For each location $\mathbf{r}$ on a facet class $\alpha$, we can define am average survival time, $\tau_\alpha(\mathbf{r};\mathcal{D})$, determined by
\begin{equation}
\tau_\alpha(\mathbf{r};\mathcal{D} )= \int_0^\infty \exp[-N_\alpha(\mathbf{r},t;\mathcal{D}) ]\,dt 
\end{equation}
where $\exp[-N_\alpha(\mathbf{r},t)]$ is the probability that site $\mathbf{r}$ is still unoccupied by time $t$. On an infinitely sized facet, the extended transformed size, $N_{\alpha}$, would not depend on the location $\mathbf{r}$, but for a finitely sized facet it is influenced by the proximity of $\mathbf{r}$ to the boundary and therefore depends on the details of the domain $\mathcal{D}$. The extended transformed size is expressed in terms of the intersection of nucleation time cone originating at $\mathbf{r}$ and the domain 
\begin{align}
    N_\alpha(\mathbf{r},t;\mathcal{D})&=
    \int_0^t
    \sum_{c}J_{\alpha,c}\big|\mathcal{C}_{\alpha}\big(\mathbf{r}, t-t';\mathcal{D}_c\big)\big|
    \,dt' \\
    \mathcal{C}_{\alpha}(\mathbf{r}, t; \mathcal{D}_c)&=\big\{\mathbf{x}\in\mathcal{D}_c \mid |\mathbf{r}-\mathbf{x}| <u_{\alpha}t\big\}
\end{align}
where $c$ indexes sites of the domain in the interior, on an edge, or at a corner, each of which have potentially distinct nucleation rates $J_{\alpha,c}$, and $u_{\alpha}$ is the step velocity of a nucleus on facet type $\alpha$, which sets the size of the unrestricted time cone. The area of the restricted time cone, $|\mathcal{C}_{
\alpha}|$, is non-dimensionalized by the lattice constant. One can equivalently think of $N_\alpha(\mathbf{r}, t)$ as the average number of nuclei that would reach site $\mathbf{r}$ before time $t$, if the nuclei did not interact with each other. The survival probability is then simply the Poisson distribution for the number of these independent nucleation events evaluated at zero.

The facet velocity is defined through an approximation to the mean layer-completion time
\begin{equation}
    V_\alpha(\mathcal{D})=\frac{a}{\max_{\mathbf{r}\in \mathcal{D}} \tau_\alpha(\mathbf{r};\mathcal{D} )}
\end{equation}
equal to the lattice constant, $a$, divided by the time it takes the slowest converting site to become covered by a new layer on average. This construction generalizes the kinetic Wulff as the velocity is not only a property of orientation $\alpha$, but also a function of the instantaneous facet size through $\mathcal{D}$. 

This kinetic theory allows for a phenomenological parameterization of site dependent nucleation rates and step velocities. However, to ground the observation we have explored it using rates deduced from a two species lattice gas model that includes both metal type sites that make up the nanocrystal, $n$, and ligand type sites that passivate the surface, $l$ \cite{oaks_leaf_mechanism_2025}. The rates to add or remove a species $k_{i,\pm}$ with $i=\{n,l\}$ are
\begin{equation}
k_{i,+} = k_0 e^{\beta \Delta \mu_i - \beta \varepsilon_i z_\alpha} \qquad k_{i,-} = k_0 e^{-\beta \varepsilon_i z_\alpha}
\end{equation}
where $\beta \Delta \mu_i$ is the chemical potential difference for moving a species from the solution to the nanoparticle in units of the thermal energy $1/\beta$, $\varepsilon_i$ is the interaction energy between like species, $z_\alpha$ the coordination number of facet type $\alpha$, and $k_0$ the bare attachment rate. For simplicity, we assume that the ligands equilibrate quickly so that they establish a mean occupation, $\rho_l$, at each facet area. Under such an assumption, we approximate the facet as switching between ligand-poor and ligand-rich states, whose free-energy difference contains an extensive binding contribution proportional to facet area $A$ and a boundary penalty proportional to perimeter $P$,
\begin{equation}
    \rho_l(A)=
    \left[
    1+\exp\left(
    -\beta\Delta\mu_l\frac{A}{a^2}
    +\beta\epsilon_l\frac{P}{2a}
    \right)
    \right]^{-1},
\end{equation}
where the area term favors ligand adsorption, while the perimeter term penalizes the unsatisfied ligand contacts. Consequently, a finite facet crosses over from weak to strong blocking at a critical area set by the ratio of ligand chemical potential to ligand--ligand cohesion. 

Due to the presence of the ligands, nucleation rates can depend on the site, as addition of metal proceeds by disrupting favorable ligand-ligand interactions. The effective attachment $k_+$ and removal $k_-$ rates for the metal in the first of the ligands become,
\begin{equation}
k_{\pm} =k_{n,\pm} \left [ (1-\rho_l) + \rho_l \frac{k_{l,\mp}}{k_{l,+}+k_{n,+}}\right ]
\end{equation}
which implies a step speed of $u_\alpha=a \left (k_+-k_-\right )$ and nucleation rates from Becker-Doring theory \cite{becker1935kinetische,limmer2024statistical},
\begin{equation}
J_{c,\alpha}=\frac{k_{+}^{(c)} u_\alpha}{a k_{-}^{(c)} +u_\alpha}    
\end{equation}
where 
\begin{equation}
\frac{k_{+}^{(c)}}{k_{n,+} } =\left [ (1-\rho_l) + \rho_l e^{-\beta \varepsilon_l z_{l,\alpha}^{(c)}} \frac{k_0}{k_{l,+}+k_{n,+}}\right ]
\end{equation}
and $k_{-}^{(c)} =e^{\beta \varepsilon_n z_\alpha} k_-$, where $c$ denotes the codimension of the nucleation site with $c=\{2,1,0\}$ corresponding to bulk, edge, and corner nucleation and $z_{l,\alpha}^{(c)}$ the corresponding number of neighbors. For a $d=2$ facet, $z_{l,100}^{(c)}=(4,3,2)$ while $z_{l,111}^{(c)}=(6,4,3)$. In this model ligands suppress nucleation in the bulk, favoring it rather at the edges or corners. 

For a bare faceted metal nanoparticle, or a small facet such that it is unlikely to be covered with ligands, the resultant size dependence of the growth velocity is weak. For such facets, nucleation will predominately occur in the interior of the facet and grow out towards the edge because there are more sites available. The survival times for two representative FCC surfaces, a square $\{100\}$ facet and a hexagonal $\{111\}$ facet,  are shown in Fig.~\ref{fig:size-dependent-growth-laws}. The resultant area dependent growth velocities are shown in Fig.~\ref{fig:size-dependent-growth-laws} and are relatively independent of facet area. 

The addition of ligands that block growth and competitively bind the surface sharpen the size dependent growth velocity substantially because ligand coverage itself depends on facet area and renders the energetics of corner, edge and bulk nucleation distinct.  Figure~\ref{fig:size-dependent-growth-laws} summarizes consequences of ligand-dependent growth on the survival time and size dependent growth velocities. The left panels show the expected completion time across representative square $\{100\}$ and hexagonal $\{111\}$ facets. At small area, ligand coverage is low
and the layer-completion pattern is close to homogeneous as there is little difference between interior, edge, and corner nucleation rates. The sites that take the longest to be transformed lie at the corners because this is where the nucleation time cone is most restricted by the boundary. For large areas, ligand blocking is strong and the boundary rates dominate over the suppressed interior rate. The slowest regions shift and the completion-time maps become controlled primarily by the propagation of nuclei at the corners toward the center of the facet. These maps are the geometric origin of a facet-area-dependent velocity. 
The right panels of Fig.~\ref{fig:size-dependent-growth-laws} show the resulting $V_\alpha(A)$ for the two facet families. As $\rho_l$ rises through its finite-facet crossover, both the step speed and the nucleation rates are suppressed \cite{liao2014facet}, producing a sharp drop in $V_\alpha(A)$. The crossover occurs at different areas for the two geometries because square and hexagonal facets have different perimeter-to-area ratios and different coordination numbers. 


\begin{figure}[t]
    \centering
    \includegraphics[width=8.5cm]{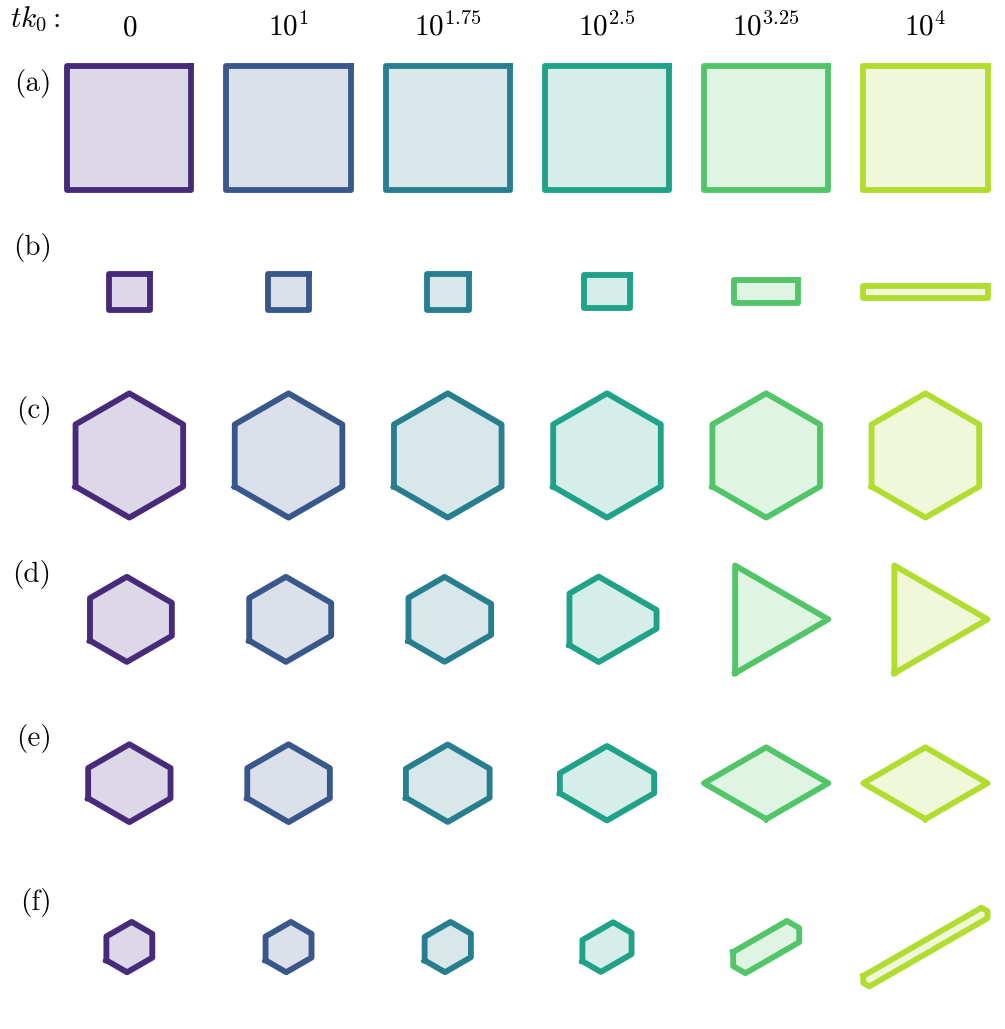}
    \caption{Area-normalized two-dimensional shape evolution under the
    size-dependent velocity law. Rows show (a) a square seed with $L_x=L_y=7a$, (b) a rectangular square-lattice seed with $L_x=6a$ and $L_y=7a$, (c) an ideal triangular-lattice hexagon, (d) a triangular seed with a single-facet perturbation, (e) a triangular seed with opposite-facet perturbations, and (f) a triangular seed with adjacent-facet perturbations. The time slices are logarithmically spaced after $t=0$, and each polygon is
    centered and rescaled to its initial area.}
    \label{fig:shape-evolution-2d-snapshots}
\end{figure}
The microscopic model establishes the possibility that under ligand mediated growth, a regime emerges where $d V_\alpha/dA<0$ so that larger facets grow more slowly than smaller ones \cite{mozaffari2019role}. To understand the impact of size dependent growth rates on morphology, we first consider square-lattices. The facet heights $h_{\pm i}$ for $i=\{x,y\}$ evolve as
\begin{align}
    \frac{dh_{\pm i}}{dt} &= V(L_i), 
\end{align}
where the growth rates depend on the linear dimension of the facet $L_x= h_{+y}+h_{-y}$ and $L_y= h_{+x}+h_{-x}$.  Growth is statistically equivalent in the $x$ and $y$ directions on a square lattice as the exposed one dimensional facets are energetically and geometrically identical. Provided an initial condition, the deterministic equations can be numerically evolved using a length-dependent velocity law obtained from the same finite-facet Avrami construction. Examples of growth trajectories are shown in Fig.~\ref{fig:shape-evolution-2d-snapshots}. For an initial condition that is square, $L_x=L_y=7a$, the deterministic equations preserve that symmetry, as shown in Fig.~\ref{fig:shape-evolution-2d-snapshots}a). However this symmetric growth is unstable and for an infinitesimal perturbation, $L_x=L_y+a$, the square lattice grows into a highly elongated rectangle, Fig.~\ref{fig:shape-evolution-2d-snapshots}b). This growth instability can be analyzed exactly leading to an identification of a pitchfork bifurcation~\cite{oaks_leaf_mechanism_2025}.  
 
For triangular-lattice seeds, six close-packed facet heights of an initial hexagon evolve as
\begin{equation}
    \frac{dh_i}{dt}=V(L_i),\qquad
    L_i=\frac{2}{\sqrt{3}}\left(h_{i+1}+h_{i-1}-h_i\right),
\end{equation}
with indices taken modulo six. Triangular-lattice calculations initialized in a regular hexagon expose equivalent one dimensional facets. However, unlike the square lattice, the vectors normal to the facets are not orthogonal. As a consequence, facets are not conserved during growth. Nevertheless initialization from a regular hexagon with side length $7a$ preserves the hexagon, as shown in Fig.~\ref{fig:shape-evolution-2d-snapshots}c). Equal-area perturbations from a regular hexagon, which include a single displaced facet, two opposite displaced facets, and two adjacent displaced facets, shown in Figs.~\ref{fig:shape-evolution-2d-snapshots}d-f) all exhibit symmetry lowering growth. In each triangular perturbation, the selected perturbed heights are scaled as $h_i\rightarrow h_i \exp(0.2)$ and  all heights are rescaled so that the initial area matches that of the regular hexagon. The modest single-facet and adjacent-facet perturbations are amplified most strongly, growing a triangular or rod shape. The opposite-facet perturbation grows a shape closer to a rhombus.

\begin{figure}[b]
    \centering
    \includegraphics[width=\linewidth]{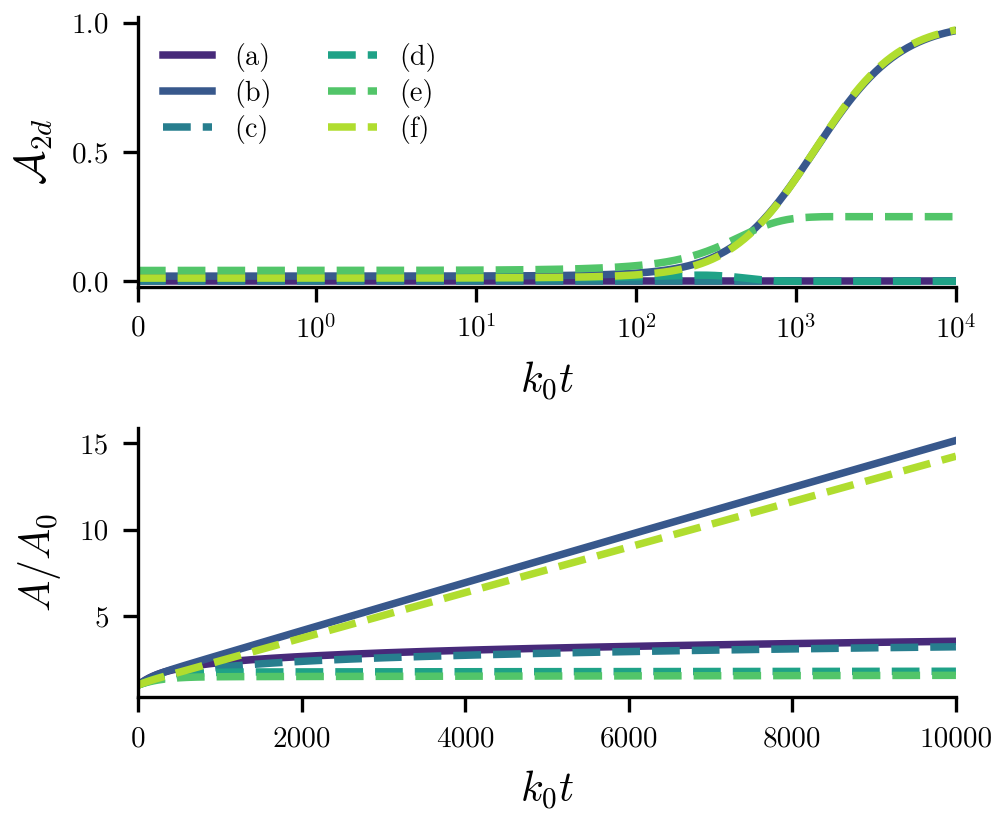}
    \caption{Shape and size trajectories for the six cases in
    Fig.~\ref{fig:shape-evolution-2d-snapshots}. The upper panel shows the
    shape asymmetry, and the lower panel shows the crystal area relative to its
    initial area. Legend labels correspond to the snapshot rows.}
    \label{fig:shape-metrics-2d-six-cases}
\end{figure}

The growth and asymmetry of the same trajectories are quantified in 
Fig.~\ref{fig:shape-metrics-2d-six-cases}. The asymmetry is quantified by computing the area-normalized second central moment
\begin{equation}
    \mathbf{G}_{2d}=\frac{1}{A}
    \int
    (\mathbf{r}-\bar{\mathbf{r}})
    (\mathbf{r}-\bar{\mathbf{r}})^{\mathsf{T}}\,dA \, ,
\end{equation}
where $\bar{\mathbf{r}}$ denotes the polygon's centroid. If $\lambda_1\leq\lambda_2$ are the eigenvalues of $\mathbf{G}_{2d}$, the shape asymmetry is
\begin{equation}
    \mathcal{A}_{2d}=
    \frac{(\lambda_2-\lambda_1)^2}{(\lambda_1+\lambda_2)^2}.
\end{equation}
which is invariant to translation and uniform rescaling, vanishes for isotropic shapes, and approaches unity for very slender objects. The area panel confirms that all six seeds grow over the same time interval, though the asymmetric shapes grow substantially more than those that preserve the symmetry of the seed.  The ideal square and hexagon remain at zero asymmetry while the initially rectangular seed, the single-facet triangular seed, and the adjacent-facet triangular seed all approach large asymmetry, demonstrating that size-dependent growth can amplify small deterministic differences into macroscopic shape anisotropy. The opposite-facet triangular perturbation grows more slowly in asymmetry because its residual twofold symmetry keeps two opposing long facets dynamically paired.

We next considered the generalized kinetic Wulff construction in three dimensions applied an FCC nanocrystal lattice. We consider seed polyhedron bounded by six $\{100\}$ and eight $\{111\}$. Each facet is specified by an outward unit normal $\hat{\mathbf{n}}_i$ and a height $h_i$, such that all points $\mathbf{r}$ in the crystal satisfy $\hat{\mathbf{n}}_i\cdot \mathbf{r}\leq h_i$ and the deterministic growth law is
\begin{equation}
    \frac{dh_i}{dt}=V_{\alpha(i)}[A_i(\mathbf{h})],
\end{equation}
where $A_i$ is the instantaneous area of facet $i$ and $\alpha(i)$ denotes whether the facet belongs to the $\{100\}$ or $\{111\}$ family. The nonequivalent facets can have distinct growth laws because of the different coordination numbers, ligand binding energies, and ligand packing densities. In the following, we consider  a symmetric reference seed formed from a cuboctahedron, with equal heights within each crystallographic family. 

Rather than employing the explicit lattice gas model to study the impact of area dependent growth velocities in three dimensions, we used an empirical growth law for each
facet family,
\begin{equation}
    V_\alpha(A)=V_{\mathrm{s},\alpha}
    +\frac{V_{\mathrm{l},\alpha}}
    {1+\exp[(A-A_{\mathrm{c},\alpha})/w]},
\end{equation}
where $\alpha\in\{\{100\},\{111\}\}$, $V_{\mathrm{l}}$ is the excess velocity of small facets, $V_{\mathrm{s}}$ is the large-facet velocity, $A_c$ is a crossover area, and $w=5a^2$. This form captures the same qualitative feature as the microscopic calculations in Fig.~\ref{fig:size-dependent-growth-laws}, small facets can grow much faster than large facets, and the crossover area can depend on facet family.

\begin{figure}[t]
    \centering
    \includegraphics[width=8.5cm]{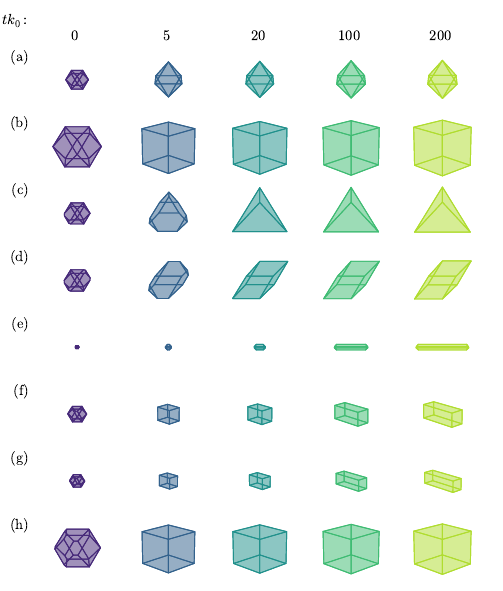}
    \caption{FCC shape evolution for the global rod comparison. Rows show
    (a) the unperturbed $\{111\}$-threshold regime, (b) the unperturbed
    $\{100\}$-threshold regime, (c) a single $\{111\}$ perturbation, (d)
    opposing $\{111\}$ perturbations, (e) adjacent $\{111\}$ perturbations,
    (f) a single $\{100\}$ perturbation, (g) opposing $\{100\}$ perturbations,
    and (h) adjacent $\{100\}$ perturbations. Time increases from left to
    right. Each polyhedron is centered and rescaled independently for
    visualization.}
    \label{fig:fcc-global-rod-snapshots}
\end{figure}

We have explored the outcomes of growth from a cuboctahedron seed with
$h_{100}=5.5a$ and $h_{111}/h_{100}=1.15$ under different initial shape perturbations and kinetic regimes. In Fig.~\ref{fig:fcc-global-rod-snapshots} we show the growth trajectories  under two swapped kinetic regimes. In the $\{111\}$-threshold regime, $A_{c,111}=120a^2$,
$A_{c,100}=1000a^2$, $V_{\mathrm{l},100}=10$, and
$V_{\mathrm{l},111}=1$. In the $\{100\}$-threshold regime,
$A_{c,100}=120a^2$, $A_{c,111}=1000a^2$, $V_{\mathrm{l},111}=10$, and
$V_{\mathrm{l},100}=1$. In all cases, $V_{\mathrm{s},100}=V_{\mathrm{s},111}
=10^{-3}$. Each regime was run from an unperturbed seed and from seeds with a height perturbation $\delta h=a$ applied to either one selected facet, two opposing or adjacent equivalent facets.

We find the unperturbed seeds remain symmetric, showing that the empirical growth law alone does not generate asymmetric shapes in the absence of a perturbation.  Perturbing a single $\{111\}$ facet grows the cuboctahedron into a symmetry lowered tetrahedron, as has been observed in gold \cite{sun2021understanding} as well as cobalt zinc ferrite~\cite{amin2026kinetically} nanoparticles. Perturbing adjacent $\{111\}$ facets in the $\{111\}$-threshold regime gives the strongest rod-like amplification, with the shape evolving toward a long object bounded by persistent $\{111\}$ facets. In the swapped $\{100\}$-threshold regime, single and opposing $\{100\}$ perturbations also produce rods, although the adjacent $\{100\}$ perturbation largely relaxes back toward a compact cubic morphology. As in two dimensions, the nonlinear growth law converts selected infinitesimal or finite seed asymmetries into persistent macroscopic anisotropy, but the outcome depends on which facet family is placed near its crossover area. 

\begin{figure}[t]
    \centering
    \includegraphics[width=\linewidth]{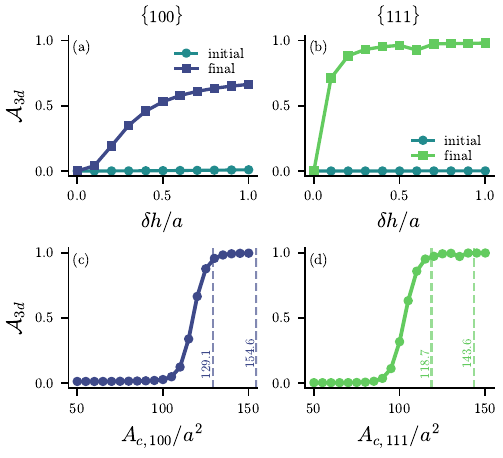}
    \caption{Sensitivity of FCC rod formation to seed perturbation and
    crossover area. The left column shows the $\{100\}$-faceted rod parameter
    set, and the right column shows the $\{111\}$-faceted rod parameter set.
    (a,b) Initial and final volume-weighted asymmetry as a function of
    perturbation height $\delta h$. (c,d) Final asymmetry as a function of the
    relevant crossover area at fixed $\delta h=a$. Dashed vertical lines mark
    the surviving facet areas at which the complementary facet family
    disappears during rod formation.}
    \label{fig:fcc-sensitivity-comparison}
\end{figure}

To understand the susceptibility of rod growth, we defined a measure of asymmetry analogous to $\mathcal{A}_{2d}$. For each convex FCC polyhedron, we compute the volume-normalized second central moment
\begin{equation}
    \mathbf{G}_{3d}=\frac{1}{V}
    \int_\Omega
    (\mathbf{r}-\bar{\mathbf{r}})
    (\mathbf{r}-\bar{\mathbf{r}})^{\mathsf{T}}\,dV 
\end{equation}
where $\bar{\mathbf{r}}$ is the volume centroid. If $\lambda_1\leq\lambda_2\leq\lambda_3$ are the eigenvalues of $\mathbf{G}_{3d}$, the 
asymmetry can be defined as
\begin{equation}
    \mathcal{A}_{3d}=
    \frac{
    (\lambda_3-\lambda_2)^2+
    (\lambda_3-\lambda_1)^2+
    (\lambda_2-\lambda_1)^2}
    {2(\lambda_1+\lambda_2+\lambda_3)^2}
\end{equation}
which is zero for an isotropic body, is unchanged by translation,
rotation, or uniform rescaling, and approaches unity for a highly elongated rod. Figure \ref{fig:fcc-sensitivity-comparison} a,b) presents the  value of $\mathcal{A}_{3d}$ at end of a long growth trajectory as a function of the amplitude of the initial asymmetry perturbation. We find that the two rod-forming regimes have different sensitivities to initial asymmetry. The $\{100\}$-faceted rod case amplifies the perturbation gradually. The final asymmetry increases smoothly over the full range of $\delta h$. The $\{111\}$-faceted rod case is much more sensitive, with a small perturbation of order $0.1a$ already producing a strongly asymmetric rod. This contrast indicates that the linearized growth dynamics near the symmetric truncated-octahedron seed is not universal. Ultimate morphology depends on which facet family is placed near the size-dependent growth crossover and on the symmetry of the perturbation.

To better understand the mechanism of rod-like growth, we studied the dependence of rod growth on crossover area, $A_{c,100}$ for the $\{100\}$-faceted rods and $A_{c,111}$ for the $\{111\}$-faceted rods. The crossover-area dependence shown in Fig. \ref{fig:fcc-sensitivity-comparison} c,d) illustrate a sharp kinetic threshold. For both rod families, the final asymmetry remains small until the crossover area approaches the facet areas present when the complementary facet family disappears. The dashed vertical lines mark those areas, for the $\{100\}$-faceted rod case, they are the surviving $\{100\}$ facet areas when the $\{111\}$ facets vanish, and for the $\{111\}$-faceted rod case they are the surviving $\{111\}$ facet areas when the $\{100\}$ facets vanish. Rod formation occurs when size-dependent kinetics keep one set of facets in the fast-growth regime long enough for small symmetry-breaking differences to be amplified before the competing facets disappear. Provided a shape can evolve from an initial perturbation to a size in which it first encounters an inflection in growth velocity, a linear stability analysis could be constructed by diagonalizing the Jacobian matrix with elements $J_{ij}=(d V_{\alpha(i)[A_i]}/dA) (d h_i/d A_j)$ to predict which shape is stable under growth. Such an analysis establishes that rod growth from a cuboctahedron occurs from quadrupolar facet-height initial perturbations, as in the opposite-facet perturbation for $\{100\}$ prevalent crystals. Quadrupolar facet-height perturbations favor both tetrahedra and rods with remaining $\{111\}$ and nonlinearities select which are ultimately formed.

The core insight of this work is that shape symmetry breaking in nanocrystal growth can be deterministic without requiring an explicitly asymmetric growth environment. Thermal fluctuations in a nanoscale seed inevitably generate small differences among symmetry-equivalent facets. In a conventional kinetic Wulff construction those differences are inconsequential and ignored \cite{boukouvala2021approaches}. Equivalent facets advance at the same rate and the construction preserves the symmetry of the velocity set. Finite facets change this picture qualitatively. If the kinetic Wulff velocity depends on the instantaneous facet length or area, then two symmetry-equivalent facets with slightly different sizes obey the same microscopic law, but move at different instantaneous speeds. The growing shape therefore feeds back on its own
kinetics, converting thermally accessible seed-shape fluctuations into macroscopic anisotropy. This feedback provides a useful way to generalize the kinetic Wulff construction. The relevant kinetic input is no longer only a velocity for each facet family, $V_\alpha$, but a growth law, $V_\alpha(A)$. 

Many established mechanisms for symmetry lowering rely on a structural or chemical fluctuation that renders one or a subset of facets statistically inequivalent, such as planar defects, differential adsorption, or heterogeneous facet reactivity \cite{gilroy2017symmetry,nguyen_colloidal_2022}.
Those microscopic ingredients may be essential insofar as they determine the finite-facet velocity functions, including the crossover areas, and the magnitude of the small- and large-facet rates. Once such an area dependence exists, unbiased thermal seed-shape fluctuations can supply the symmetry-breaking perturbation. The experimentally important controls are therefore the finite-facet kinetic crossovers as much as the asymptotic large-particle velocity ratios which are dependent on ligand coverage, nucleation barriers, step propagation, and edge or corner reactivity.

The calculations here are deterministic. As a consequence they do not predict the probability of specific shapes, as they would result from thermally accessible seed fluctuations or stochastic nucleation histories \cite{ray2018importance}. Models employed are also phenomenological, meant to expose the amplification mechanism over a broad parameter range, rather than deriving the velocity curve for a specific molecular system. These approximations are deliberate. They separate the question of dynamical stability from the microscopic origin of the rate law. A predictive theory of product yields will need to combine both pieces by sampling seed-shape fluctuations, computing facet-family velocity functions from microscopic models or deducing them from experiment, and propagating the resulting ensemble through the stochastic version of the evolution equations developed here.

\emph{Acknowledgments} This work was supported by the U.S. Department of Energy, Office of Science, Office of Basic Energy Sciences, Materials Sciences and Engineering Division under Contract No. DE-AC02-05-CH11231 within the in-situ TEM program (KC22ZH). Codex 5.6 Sol was used in manuscript preparation, the writing of simulation code, and the analysis of data, all with  supervision and explicit verification. 

\emph{Data and code availability} The simulation code, analysis and plotting script, are available on zenodo, https://zenodo.org/records/22032024. 

\bibliographystyle{apsrev4-2}
\bibliography{references}

\end{document}